\documentclass[fleqn,usenatbib]{mnras}

\usepackage[utf8]{inputenc}
\usepackage[T1]{fontenc}
\usepackage{newtxtext,newtxmath}
\usepackage{graphicx}
\usepackage{amsmath}
\usepackage{multirow}
\usepackage{hyperref}
\usepackage{bookmark}
\usepackage[most]{tcolorbox}
\usepackage{rotating}
\usepackage{ulem}
\usepackage{longtable}
\usepackage{lscape}
\usepackage{caption}
\usepackage{bm}
\usepackage{cleveref}
\usepackage{float}
\usepackage{type1cm}

\title[Infrared-Excess in VVV]{Infrared-excess sources in VVV: candidate debris discs, SED characterization, and variability analysis}

\author[P.~Esteves et al.]{%
P.~Esteves$^{1}$,
R.~K.~Saito$^{1}$,
C.~Chavero$^{2,3}$,
V.~Fermiano$^{4,1}$,
D.~Minniti$^{5,6}$,
B.~W.~Borges$^{1,7}$,
\newauthor
C.~Cáceres$^{5}$,
Z.~Guo$^{4,8}$,
V.~D.~Ivanov$^{9}$,
R.~G.~Kurtev$^{4,10}$,
D.~Quispe$^{1}$,
R.~M.~Torres$^{1,11}$
\\\\
$^{1}$Departamento de Física, Universidade Federal de Santa Catarina, Trindade 88040-900, Florianópolis, Brazil\\
$^{2}$Observatorio Astronómico de Córdoba, Universidad Nacional de Córdoba, Laprida 854, 5000 Córdoba, Argentina \\
$^{3}$Consejo Nacional de Investigaciones Científicas y Técnicas (CONICET), Godoy Cruz 2290, Ciudad Autónoma de Buenos Aires, Argentina\\
$^{4}$Instituto de Física y Astronomía, Universidad de Valparaíso, Av. Gran Bretaña 1111, Playa Ancha, Casilla 5030, Chile\\
$^{5}$Instituto de Astrofísica, Departamento de Física y Astronomía, Facultad de Ciencias Exactas, Universidad Andres Bello, Av. Fernández Concha 700, Santiago, Chile\\
$^{6}$Vatican Observatory, Specola Vaticana, V-00120, Vatican City State\\
$^{7}$Coordenadoria Especial de Física, Química e Matemática, Universidade Federal de Santa Catarina, Jardim das Avenidas 88906-072, Araranguá, Brazil\\
$^{8}$Chinese Academy of Sciences South America Center for Astronomy (CASSACA), National Astronomical Observatories, CAS, Beijing 100101, China\\
$^{9}$European Southern Observatory, Karl Schwarzschildstr 2, 85748 Garching bei München, Germany\\
$^{10}$Millennium Institute of Astrophysics, Nuncio Monseñor Sotero Sanz 100, Of. 104, Providencia, Santiago, Chile\\
$^{11}$Instituto Federal Catarinense, 89703-720, Concórdia, Brazil\\
}
\pubyear{2026}

\date{Accepted 2026 June 08. Received 2026 June 03; in original form 2026 April 30}

\begin{document}

\maketitle
\begin{abstract}

Debris disks around main-sequence stars provide important insights into planetary formation and evolution; however,
identifying these systems in the Galactic plane, where high interstellar extinction and source crowding are significant,
remains challenging. Deep, wide-area surveys such as the VISTA Variables in the Via Lactea (VVV) ESO Public Survey provide
multi-epoch infrared observations, enabling systematic searches in these environments. We present a method to construct a
refined sample of main-sequence candidates by cross-matching data from \textit{Gaia} DR3, GLIMPSE, DECaPS, and VIRAC2, to identify
   infrared excesses through spectral energy distribution (SED) analysis, and to analyse time-series data to detect variability
potentially associated with circumstellar material or other astrophysical phenomena. Although infrared excess is commonly
associated with debris disks, contributions from unresolved binaries or line-of-sight contamination cannot be excluded. Applying this methodology to the VVV tile d077 using the VIRAC2 catalogue, we analyse an initial sample of 879,556
sources and identify 446 sources flagged by the VO SED Analyser (VOSA) excess-detection algorithm as having infrared excess, of which 171 show candidate periodic variability in the
VIRAC2 $K_{\rm s}$-band light curves. We further perform SED fitting using theoretical models to derive physical parameters
and approximate spectral types for these sources. This work provides a catalogue of infrared-excess candidates in a Galactic field affected by high extinction,
along with a reproducible method suitable for future large-scale applications.

\end{abstract}


\begin{figure*}
    \centering
    \includegraphics[width=2\columnwidth]{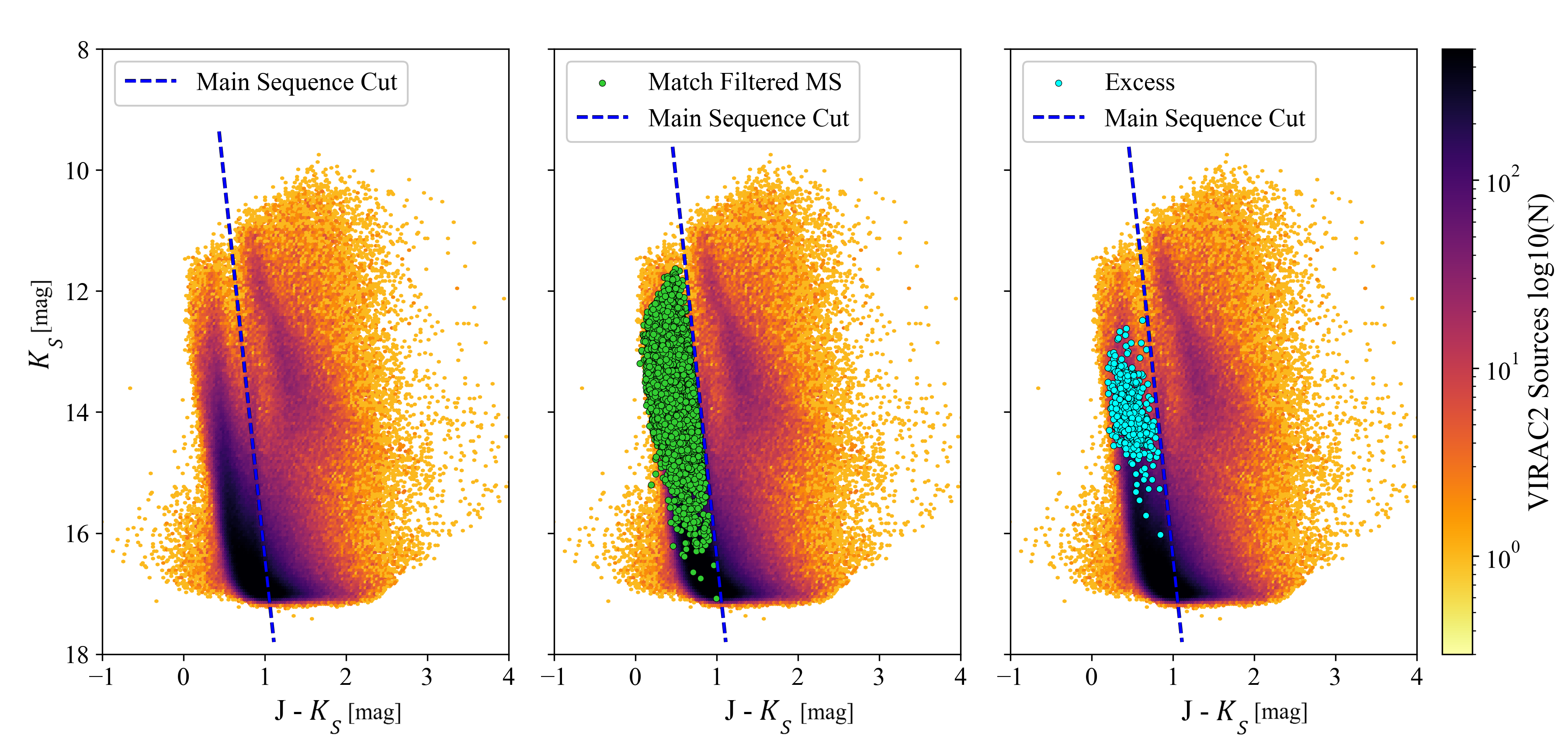}
    \caption{Colour--magnitude diagrams ($(J - K_s)$ vs.\ $K_s$) illustrating the progressive filtering of the sample. In all three panels, the solid line marks an empirical boundary used to isolate the main sequence: it is parameterised as $x = x_0 + \mathrm{slope}\,(y - y_0)$, with $(x_0, y_0, \mathrm{slope}) = (0.97, 16.0, 0.08)$. Sources located to the left of this line were classified as main-sequence candidates. \textbf{Left:} all VIRAC2 sources in the field, highlighting the main sequence and red-giant populations. \textbf{Centre:} cross-matched Gaia DR3 + VIRAC2 + GLIMPSE + DECaPS sample, where the empirical boundary isolates 12,940 main-sequence candidates. \textbf{Right:} the final 446 sources flagged as having infrared excess after all selection criteria are applied.}
    \label{fig:cmd}
\end{figure*}

\begin{figure}
    \centering
    \includegraphics[width=1\linewidth]{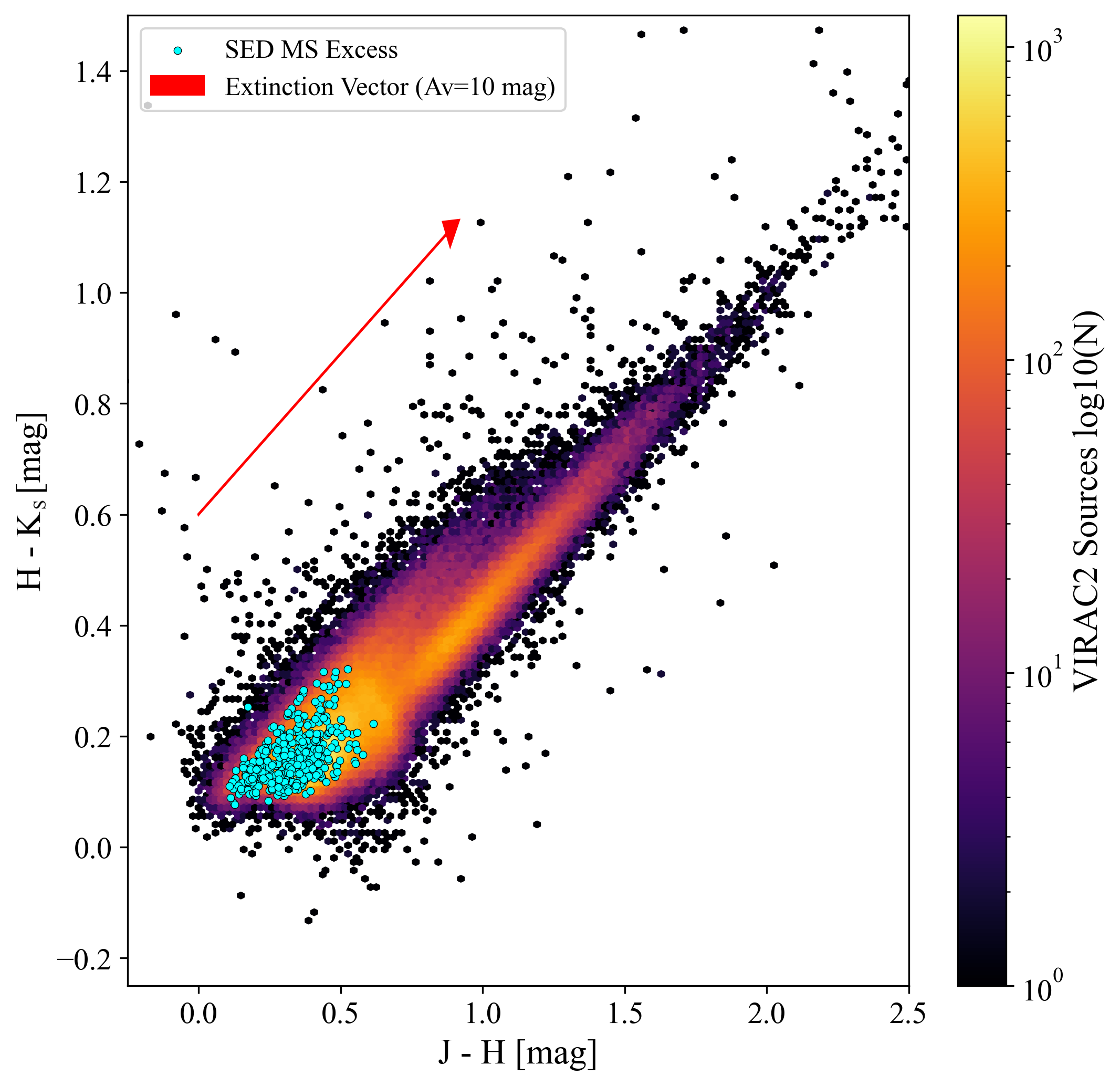}
    \caption{Colour--colour diagram $(J-H)$ versus $(H-K_s)$ for the final sample. The background density shows the distribution of VIRAC2 sources in the field, while the cyan points indicate the 446 sources in the final infrared-excess sample. The red arrow shows the theoretical extinction vector corresponding to $A_V = 10$ mag, derived from the \citealt{Fitzpatrick1999} extinction law, which is the same prescription adopted by \textit{Gaia} DR3 to estimate $A_0$. The diagram illustrates the location of the selected sources in colour--colour space relative to the expected reddening behaviour.}
    \label{fig:colorcolor}
\end{figure}

\begin{figure}
    \centering
    \includegraphics[width=1\linewidth]{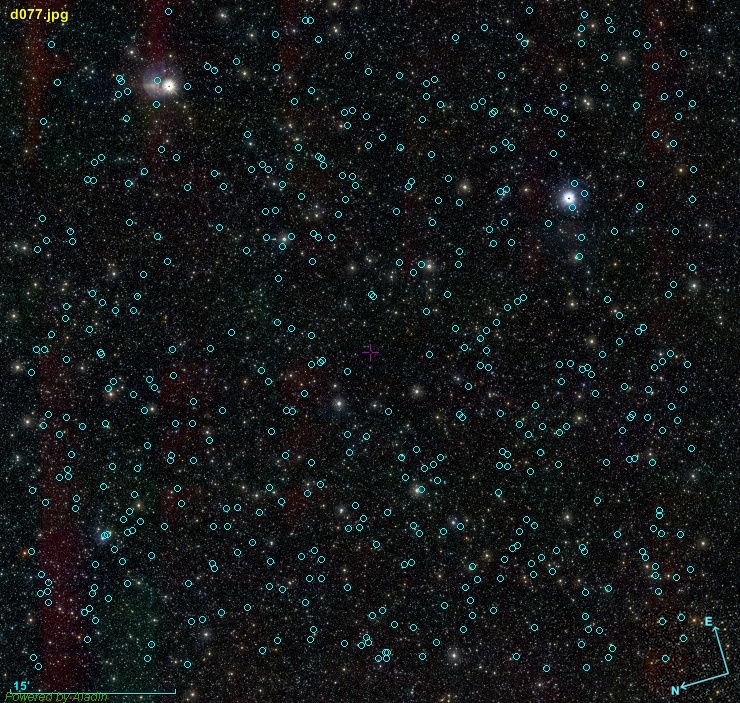}

    \caption{Spatial distribution of the 446 infrared-excess sources selected within the DECaPS footprint, overlaid on the VVV image of tile d077, centred at Galactic coordinates $(\ell, b) = (295.5^\circ,\,0.5^\circ)$ and equatorial coordinates $(\mathrm{RA},\,\mathrm{Dec}) = (11^{\mathrm h}48^{\mathrm m}36.32^{\mathrm s},\,-61^\circ27'42.5'')$. The displayed field corresponds to a $1.0\,\mathrm{deg}^2$ region, matching the area adopted from the DECaPS survey. The blue markers indicate the projected positions of the selected sources in equatorial coordinates. The arrows labelled N and E indicate the directions of increasing declination (north) and right ascension (east) in the ICRS reference frame.}
   \label{fig:spatial_distribution}
\end{figure}

\begin{figure*}
    \centering
    \includegraphics[width=1\textwidth]{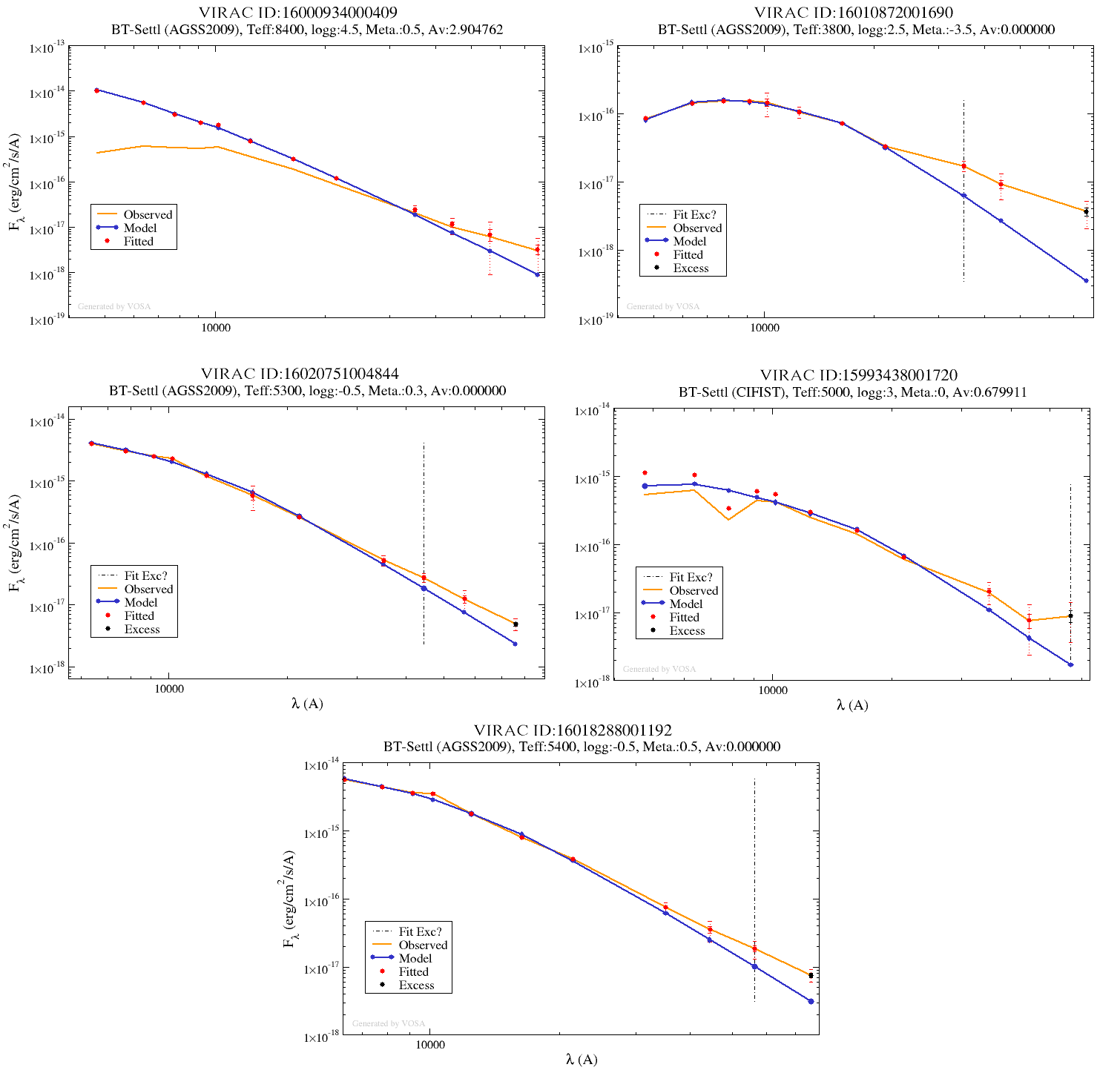}

    \caption{Observed spectral energy distributions (SEDs) for five representative infrared-excess sources constructed using photometry from DECaPS (optical), VIRAC2 (near-infrared), and GLIMPSE (mid-infrared). Red points correspond to the observed photometric fluxes, with vertical error bars indicating their uncertainties. The orange line represents the observed SED interpolated between the data points and normalised using Gaia photometry. The blue line shows the best-fit stellar photospheric model obtained with VOSA. The vertical dashed line marks the wavelength beyond which infrared-excess is identified. Black points indicate photometric measurements flagged as affected by excess emission and not used to constrain the photospheric fit. The parameters displayed in each panel correspond to the best-fit stellar properties derived by VOSA, including effective temperature ($T_{\rm eff}$), surface gravity ($\log g$), and extinction ($A_V$).}
    \label{fig:sed}
\end{figure*}

\begin{figure*}
    \centering

    \includegraphics[width=1\textwidth]{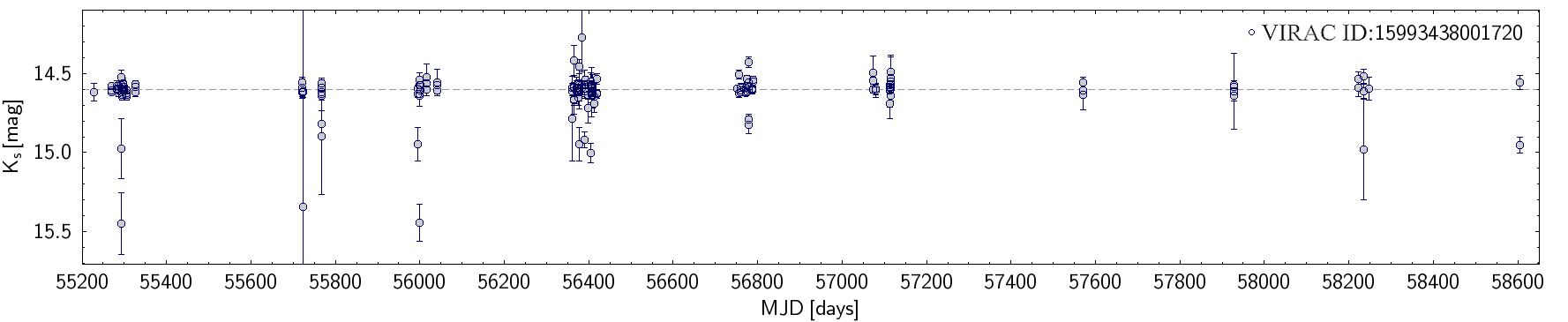}

    \includegraphics[width=1\textwidth]{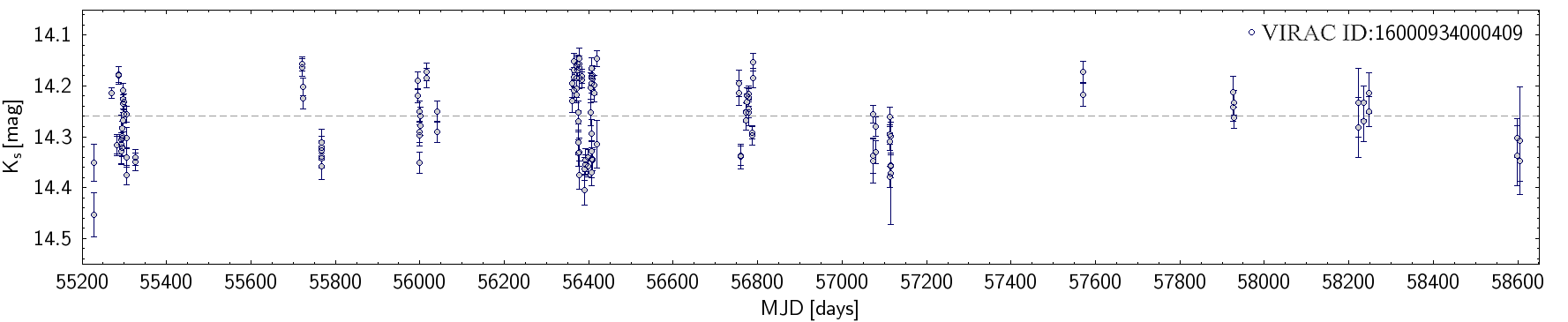}

    \includegraphics[width=1\textwidth]{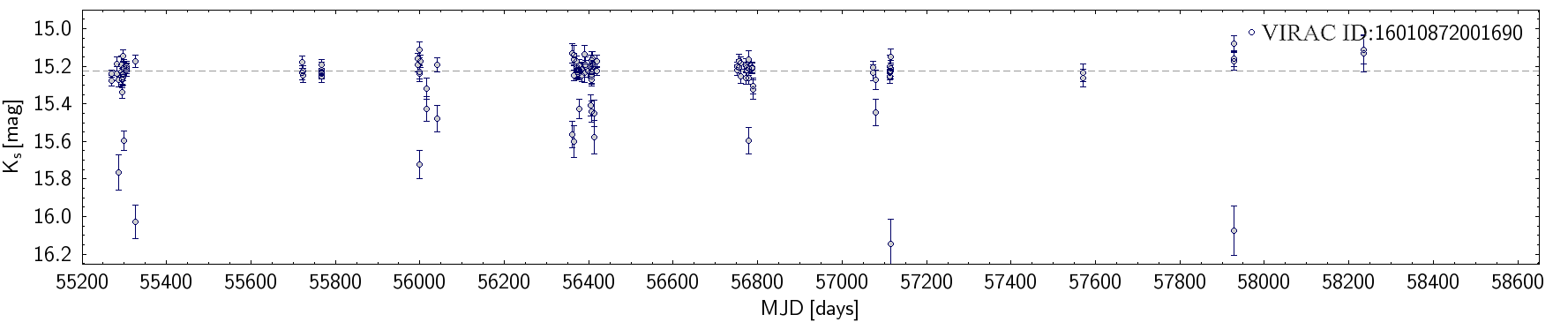}

    \includegraphics[width=1\textwidth]{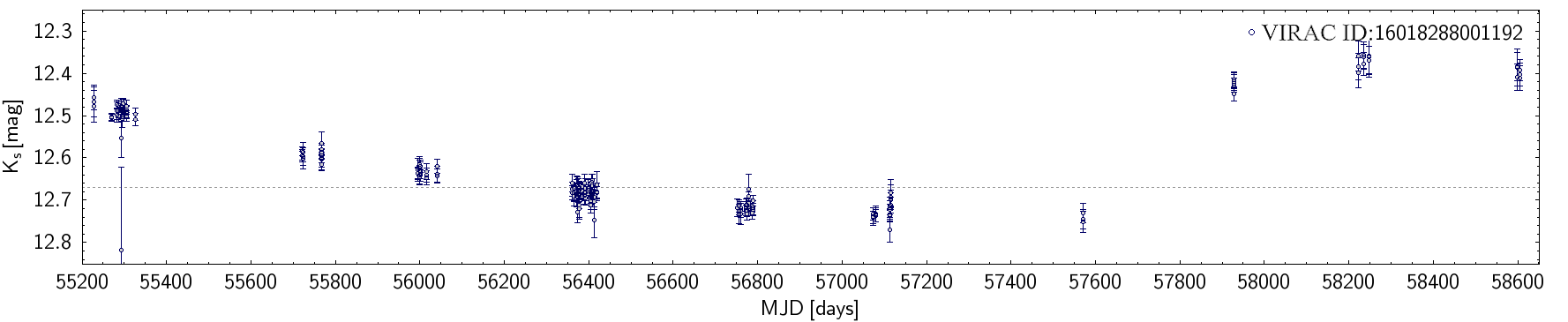}

    \includegraphics[width=1\textwidth]{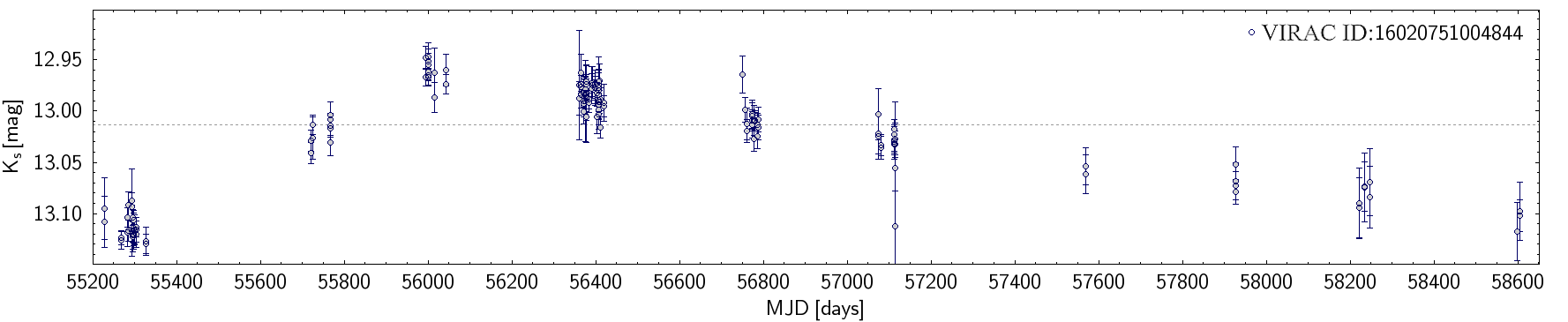}

    \caption{$K_{\rm s}$-band light curves of five representative infrared-excess sources identified in this work, corresponding to the same sources shown in Fig.~\ref{fig:sed}. Each panel corresponds to a different source, identified by its VIRAC2 ID indicated within the panel. Points represent individual epoch measurements from VIRAC2. These light curves complement the SED analysis by revealing variability that may be associated with circumstellar material or intrinsic stellar activity.}
    \label{fig:lightcurve}

\end{figure*}

\begin{figure}
    \centering

    \includegraphics[width=0.5\textwidth]{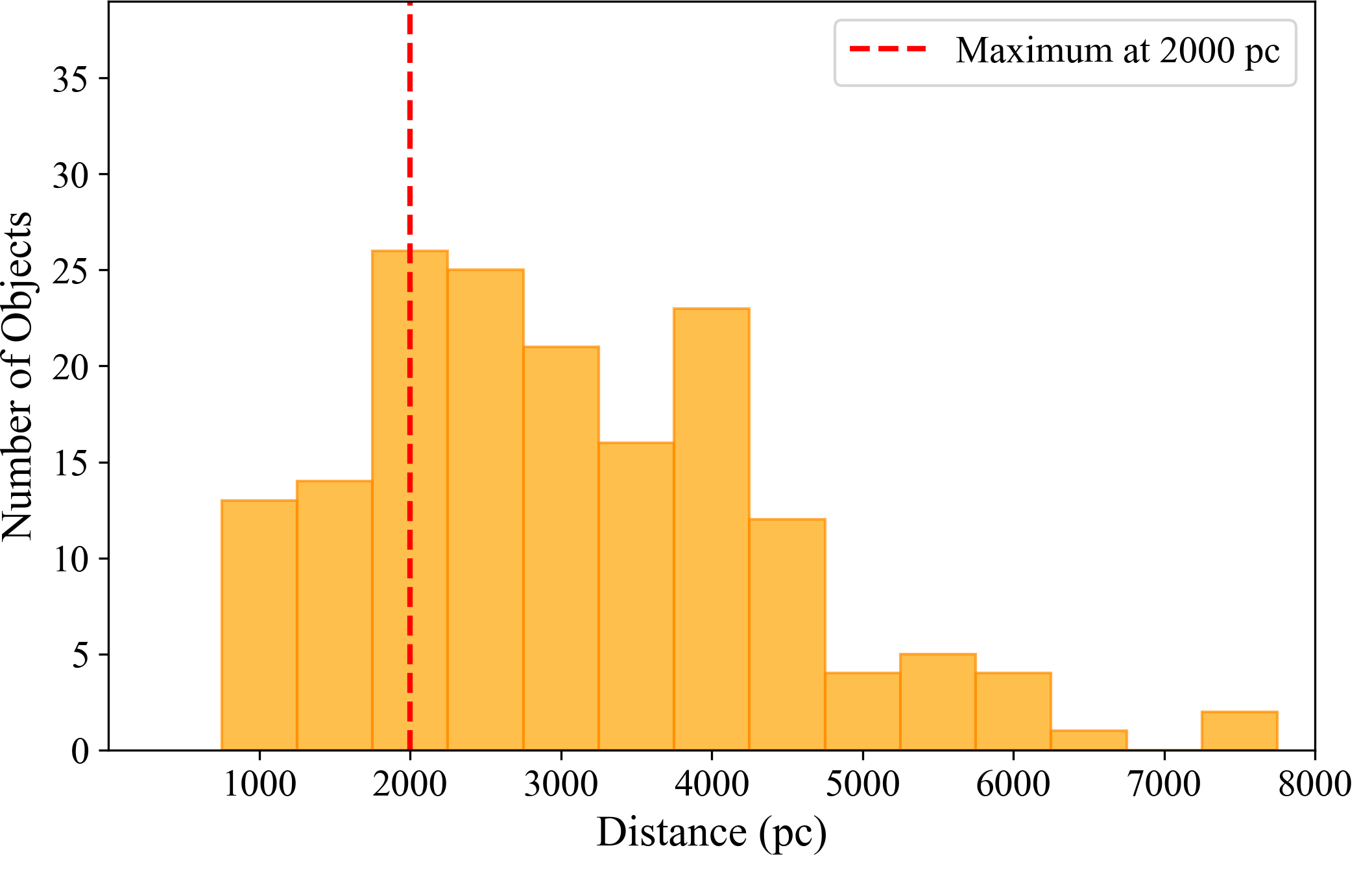}
    \caption{Distribution of distances for the 171 infrared-excess variable sources identified in this work. Distances are derived from Gaia DR3 parallaxes when available. The distribution exhibits a maximum around 2000 pc, as highlighted by the red dashed line.}
    \label{fig:distance}
\end{figure}

\begin{figure}
    \centering
    \includegraphics[width=0.5\textwidth]{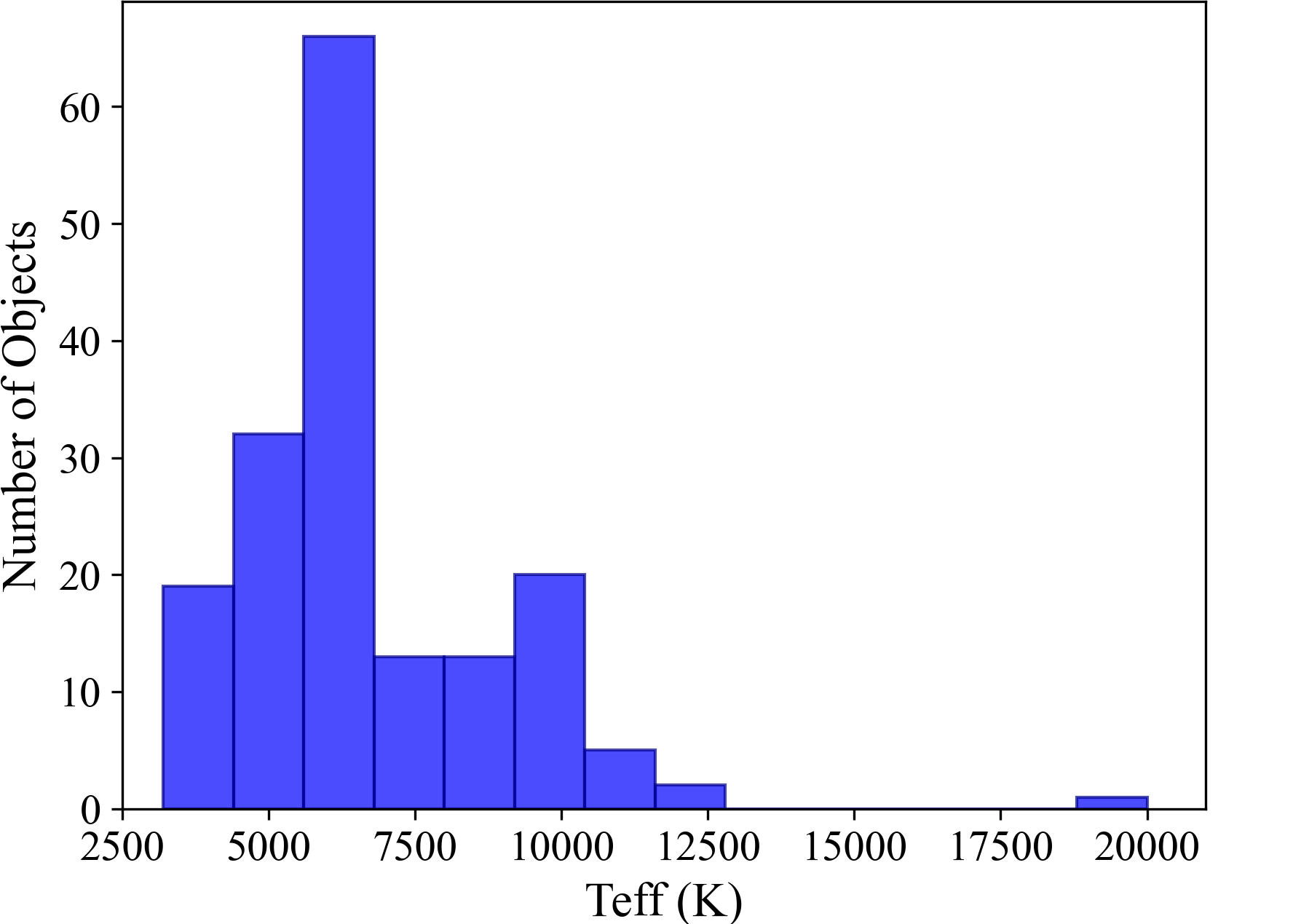}
    \caption{Distribution of effective temperatures for the 171 infrared-excess variable sources. The distribution reflects the range of stellar types included in the sample and is derived from SED fitting using VOSA.}
    \label{fig:teff}
\end{figure}

\section{Introduction}

Understanding the formation and evolution of planetary systems remains a central goal of modern astrophysics. While the number of confirmed exoplanets has grown dramatically in recent decades, circumstellar debris disks provide an essential complementary window into planetary architectures and dynamical histories \citep{Lissauer1993, Wyatt2008, Hughes2018}. These second–generation disks, composed of dust produced by planetesimal collisions, exhibit infrared (IR) excesses that encode information on dust composition, disk geometry, and evolutionary stage.

Debris disks differ fundamentally from the gas-rich primordial disks observed around young stellar sources: their IR signatures are typically weaker, and their evolution is shaped primarily by collisional cascades, dynamical stirring, and radiative forces \citep{DAlessio1999, Wyatt2008}. Large-area IR surveys — from IRAS to \textit{Spitzer}, WISE, and \textit{Herschel} — have revealed hundreds of debris-disk systems and provided key demographic insights. However, these surveys tend to avoid the highly extincted and crowded regions of the Galactic plane, where variable reddening, source confusion, and contamination by evolved stars compromise the identification of genuine IR excesses \citep{Aumann1984, Su2006, Eiroa2013}. As a consequence, debris disks in these environments remain systematically underrepresented.

Previous searches for mid-infrared excess sources in the Galactic plane have also explored the potential of \textit{Spitzer}/GLIMPSE data to identify objects with circumstellar dust emission. For example, \citet{Chen2005}, \citet{Uzpen2007}, \citet{Hales2009}, and \citet{Maldonado2017} used mid-infrared photometry to select and characterise infrared-excess sources, including debris-disk candidates and other dusty stellar populations. These studies demonstrated the value of GLIMPSE for identifying warm dust emission in crowded Galactic fields, while also highlighting the challenges posed by source confusion, extinction, and contamination. The present work builds on this approach by combining GLIMPSE with DECaPS, \textit{Gaia} DR3, and VIRAC2 photometry, and by adding near-infrared time-domain information from VIRAC2 to investigate variability among the selected infrared-excess sources.

The growing availability of deep, multi-wavelength surveys now enables more reliable searches in the Galactic plane. Optical photometry from DECaPS \citep{Schlafly2018} constrains the short-wavelength portion of the spectral energy distribution (SED), while VIRAC2, based on VVV and VVVX observations, provides near-IR measurements and multi-epoch $K_{\rm s}$-band light curves \citep{Minniti2010, Saito2012, Smith2021}. Mid-IR data from GLIMPSE \citep{Benjamin2003, Churchwell2009} are particularly sensitive to warm circumstellar dust, allowing for the detection of excess emission beyond the stellar photosphere. When combined, these datasets enable systematic SED construction and help distinguish reddening effects from genuine infrared-excess emission. Despite these advances, a systematic and homogeneous search for infrared-excess sources in the Galactic plane, combining multi-wavelength SED analysis with time-domain variability diagnostics, remains largely unexplored.

Time-domain information adds an additional diagnostic for circumstellar environments. Variability in the near-IR may arise from transient dust production, occultations by disk structures, or changes in thermal re-emission. The multi-epoch VIRAC2 light curves, therefore, provide an opportunity to identify systems whose variability is plausibly connected to circumstellar material, including disk-related eclipses or occultations by dusty clumps.

In this work, we develop a method to (i) construct a refined sample of main-sequence candidates via a multi-survey cross-match between DECaPS, VIRAC2, GLIMPSE, and \textit{Gaia} (Sections~\ref{sec:data}–\ref{sec:selection}); (ii) perform SED fitting to identify sources flagged by the VOSA excess-detection algorithm as having IR excess (Section~\ref{sec:sedmodel}); and (iii) analyse multi-epoch VIRAC2 $K_{\rm s}$-band light curves to search for periodic or coherent variability possibly associated with circumstellar dust structures (Section~\ref{sec:var}). Applying this method to the VVV tile d077, we identify 446 infrared-excess sources out of 879,556 initial sources, of which 171 show candidate periodic variability in the VIRAC2 $K_{\rm s}$-band light curves.
The main product of this work is a catalogue of infrared-excess candidate sources in a highly reddened Galactic-plane field, together with a reproducible method that can be extended to other VVV/VIRAC2 regions.
The resulting sample and derived physical parameters are presented in Section~\ref{sec:results}, followed by a discussion of their implications in Section~\ref{sec:discussion}.

\section{Data and Surveys}
\label{sec:data}

The \textit{VISTA Variables in the Via Lactea} survey (VVV; \citealt{Minniti2010,Saito2012}) is one of the most extensive near-infrared mappings of the Milky Way, covering the bulge and the southern inner disk. Observations were conducted with the 4.1-m VISTA telescope at Paranal Observatory using the VIRCAM instrument, which provides a $1.65\ \mathrm{deg}^2$ field of view with a pixel scale of $0.34''$. The survey includes imaging in the $Z$, $Y$, $J$, $H$, and $K_{\rm s}$ bands, reaching depths up to $\sim$2 magnitudes deeper than 2MASS, and delivers multi-epoch $K_{\rm s}$ photometry ideal for variability studies. Although the VVV survey includes $Z$-band observations, this band was not used in the present analysis due to its limited sensitivity in regions of high extinction and its lower completeness compared to the near-infrared bands.

The d077 tile has central coordinates $(\ell, b) = (295.346^\circ, 0.443^\circ)$, placing it directly on the Galactic plane. While a typical VVV/VVVX tile covers approximately $1.5\,\mathrm{deg}^2$, we restrict our analysis to a $1.0\,\mathrm{deg}^2$ region corresponding to the DECaPS footprint, centred at $(\ell, b) = (295.5^\circ, 0.5^\circ)$. This choice simplifies the evaluation of target density and ensures a consistent spatial coverage across all datasets used in this work.

For this work, we make use of the $Y$, $J$, $H$, and $K_{\rm s}$ single-epoch photometry provided by the VISTA Infrared Astrometric catalogue (VIRAC2; \citealt{Smith2021}), which delivers homogenised astrometric and photometric measurements over the original VVV footprint. Although spatially limited to the VVV area, it incorporates additional $K_{\rm s}$-band epochs from the VVV eXtended survey (VVVX; \citealt{Saito2024}), providing a time baseline of $\sim$14 years for the variability analysis. Together, these datasets constitute the near-infrared backbone of our SEDs and enable an independent investigation of photometric variability potentially associated with circumstellar material.

To anchor the short-wavelength regime of the SEDs, we incorporate optical photometry from the DECam Plane Survey (DECaPS; \citealt{Schlafly2018}), which provides deep $g$, $r$, $i$, $z$, and $Y$ observations over the southern Galactic plane. Mid-infrared photometry is obtained from the GLIMPSE survey \citep{Benjamin2003,Churchwell2009}, which offers coverage in the four \textit{Spitzer}/IRAC bands and is particularly sensitive to warm circumstellar dust responsible for the infrared-excess.

Astrometric information, distances, and extinction estimates are taken from \textit{Gaia} DR3 \citep{GaiaDR3}. We adopt the DR3 extinction parameter $A_0$ based on the \citealt{Fitzpatrick1999} extinction law as a first-order correction for interstellar reddening, along with geometric distances whenever available, in order to estimate intrinsic stellar properties for the SED analysis.

\section{Sample Selection}
\label{sec:selection}

To construct the working sample for this study, we combined the optical, near-infrared, and mid-infrared photometric datasets described in Section~\ref{sec:data}. Because the target field lies in a densely populated region of the Galactic disk, particular care was taken to minimise spurious associations while retaining genuine multi-wavelength counterparts. All catalogue combinations were carried out using the STILTS toolkit \citep{Taylor2006}, and the adopted matching strategy was evaluated by comparing restrictive and permissive positional matches, allowing us to identify duplicated or ambiguous associations before defining the final catalogue.

The cross-matching procedure was performed in successive stages to identify and eliminate uncertain counterparts. We first generated a restrictive match between VIRAC2 and GLIMPSE, retaining only the nearest positional counterpart within a $1''$ radius (“Match~1”). A second, more permissive pass (“Match~2”) included all potential associations out to $5\arcsec$, providing a diagnostic set to identify duplicated or ambiguous matches. By comparing these two outputs, we removed inconsistent identifications and constructed a refined multi-wavelength catalogue. This optimised sample was then cross-matched with DECaPS and \textit{Gaia}~DR3, again adopting the nearest-neighbour association within $1''$ to ensure reliable optical and astrometric counterparts.

No explicit filtering based on the \textit{Gaia}~DR3 Renormalised Unit Weight Error (RUWE) parameter was applied, in order to avoid imposing an additional astrometric selection on this crowded and highly reddened field. As a consequence, unresolved binary systems cannot be excluded based on astrometric criteria and may contribute to the observed infrared-excess in some cases.

With this consolidated catalogue, we proceeded to isolate main-sequence candidates, which define the baseline population for identifying mid-infrared excesses. The primary main-sequence selection was based on the colour--magnitude diagram rather than on the colour--colour diagram; the latter is used as an auxiliary diagnostic of reddening behaviour and consistency with the adopted extinction correction. A $(J-K_s)$ versus $K_s$ colour--magnitude diagram (Fig.~\ref{fig:cmd}) constructed from the cross-matched sources reveals a well-defined main-sequence locus. To suppress contamination from giants, highly reddened sources, and other non-main-sequence populations, we defined an empirical linear boundary along the blue edge of the distribution. The parameters defining this boundary are indicated in Fig.~\ref{fig:cmd}, ensuring reproducibility of the selection. This boundary was determined directly from the observed stellar locus and applied uniformly across the sample. Sources located blueward of this line were retained as main-sequence candidates. This approach prioritises sample purity over completeness, reducing the likelihood of false-positive IR-excess detections arising from inappropriate stellar templates, unresolved binaries, or source confusion.

Because the selection is empirical and based on broad-band photometry, some contamination by non-main-sequence populations may remain. In particular, reddened giants, evolved stars, young stellar objects, and unresolved blends may occupy regions of the colour--magnitude diagram that partially overlap with the adopted main-sequence locus. Therefore, the selection should be interpreted as a photometrically refined main-sequence candidate sample rather than as a spectroscopically confirmed dwarf-star sample. The subsequent SED fitting provides an additional consistency check on the selected sources, but it cannot fully remove degeneracies among stellar parameters, extinction, and distance.

Finally, we required sufficient photometric coverage to construct reliable spectral energy distributions. Because SED fitting depends on constraints across the optical, near-infrared, and mid-infrared regimes, we required that each source be detected in at least two bands within each contributing survey. This criterion ensures adequate sampling of the stellar photosphere and minimises degeneracies in the model fitting. Only sources satisfying this photometric completeness requirement were retained for the subsequent SED modelling and IR-excess analysis described in Section~\ref{sec:sedmodel}.

\section{SED Construction and Modelling}
\label{sec:sedmodel}

With the filtered sample defined in Section~\ref{sec:selection}, we constructed spectral energy distributions (SEDs) by combining the optical, near-infrared, and mid-infrared photometric measurements from DECaPS, VIRAC2, and GLIMPSE. The SED analysis was carried out using the \textit{Virtual Observatory SED analyser} (VOSA; \citealt{Bayo2008}), which provides a homogeneous framework for fitting multi-band photometry to grids of stellar atmospheric models.

For each source, VOSA performed automated fits to the Kurucz and BT-Settl stellar model grids, incorporating the photometric uncertainties and adopting extinction corrections based on the \textit{Gaia}~DR3 extinction parameter $A_{0}$ together with the \citet{Fitzpatrick1999} reddening law. The consistency of these extinction corrections was qualitatively assessed using the colour--colour diagram shown in Fig.~\ref{fig:colorcolor}. Best-fit parameters were obtained through $\chi^{2}$ minimisation, and the resulting model was used as the reference photosphere against which deviations at longer wavelengths were assessed.

We note that degeneracies between $T_{\rm eff}$, extinction, and distance may affect the SED fitting when using broad-band photometry alone. These degeneracies are particularly relevant in crowded and highly reddened Galactic-plane fields, where a reddened giant or blended source may reproduce colours similar to those of a main-sequence star under a different combination of extinction and distance. The multi-wavelength coverage adopted here, together with the imposed photometric completeness criteria, helps reduce the occurrence of poorly constrained or discontinuous fits, but it does not provide a unique stellar classification for every source. For this reason, the best-fitting SED parameters are used as a consistency check rather than as a definitive spectroscopic classification.

The excess search was applied to the longest available near- and mid-infrared photometric points, particularly in the GLIMPSE/IRAC bands, where deviations from the stellar photosphere are expected to be most evident. VOSA implements an excess-detection algorithm based on the method introduced by \citet{Lada2006}, which evaluates the slope of the logarithmic relation between $\nu F_{\nu}$ and $\nu$ at infrared wavelengths. In this work, infrared-excess candidates are defined as sources flagged by the VOSA weighted-slope criterion, rather than by imposing an additional fixed signal-to-noise threshold above the photospheric model. This approach provides a systematic diagnostic for identifying deviations from purely photospheric emission, which are commonly interpreted as signatures of circumstellar dust. However, the VOSA flag should be interpreted as a candidate-selection criterion: contributions from unresolved binary companions or line-of-sight blends cannot be excluded and may also produce similar infrared-excess signatures.

To mitigate contamination by image artefacts or nearby bright sources, all infrared-excess sources were visually inspected on the VVV $K_{\rm s}$ images as a quality-control step. Sources with clear visual evidence of saturation spikes, close blends, or contamination by nearby bright sources were rejected. Ambiguous cases were conservatively removed from the final excess list. This visual inspection was not used to define the infrared excess itself, but only to reject likely spurious cases caused by image artefacts, saturation, or blending.

\section{Variability Analysis}
\label{sec:var}

To investigate the time-domain behaviour of the infrared-excess sources, we analysed their $K_{\rm s}$-band light curves using the multi-epoch photometry provided by VIRAC2. The VIRAC2 time series spans several observing seasons with irregular cadence and typically provides on the order of tens of epochs per source, sufficient to probe variability on timescales from days to years \citep{Saito2024}.

The variability analysis was initially based on the Stetson index \citep{Stetson1996}, which quantifies correlated deviations between paired measurements and is widely used to distinguish genuine variables from noise. A threshold of $S > 0.5$ was adopted to identify candidate variable sources, resulting in 356 objects exhibiting significant variability.

Periodicity was then assessed using two complementary techniques: the Lomb–Scargle periodogram \citep{Lomb1976,Scargle1982}, which is sensitive to sinusoidal or quasi-sinusoidal signals in unevenly sampled data, and the Box Least Squares (BLS) algorithm \citep{Kovacs2002}, which is optimised for detecting transit- or eclipse-like events with non-sinusoidal profiles.

We adopted a false-alarm probability threshold of $\mathrm{FAP} < 0.01$ to identify statistically significant periodic signals. Applying this criterion reduced the sample to 171 sources with candidate periodic variability.

For each of these sources, the VIRAC2 $K_{\rm s}$-band light curves were first examined in the time domain to assess variability patterns and data quality. Both Lomb--Scargle and BLS periods are reported in the final catalogue. The final periodic classification was assigned when the Lomb--Scargle signal satisfied the adopted FAP criterion, with BLS used as a complementary diagnostic for eclipse-like variability, and the phased light curve passed visual inspection. Phase-folded light curves were then constructed to evaluate the coherence and robustness of the detected periodic signals.

This procedure yields a conservative sample of variable infrared-excess sources, suitable for investigating possible links between infrared excess, circumstellar material, and intrinsic or binary-related stellar variability. The numerical results of this analysis are presented in Section~\ref{sec:results}.

\section{Results}
\label{sec:results}

The VIRAC2 catalogue for the VVV tile d077 contains 879,556 sources. Using the cross-matching procedure described in Section~\ref{sec:selection}, we obtained a sample of 36,253 sources with uniform optical-to-mid-infrared coverage.

Application of the main-sequence selection criteria outlined in Section~\ref{sec:selection} reduced this sample to 12,940 sources. Additional photometric completeness requirements yielded a final set of 3,169 sources with sufficiently sampled spectral energy distributions (SEDs).

Among these, 446 sources were flagged as having infrared excess according to the SED-fitting procedure described in Section~\ref{sec:sedmodel}. After removing image contaminants through the visual inspection described in Section~\ref{sec:sedmodel}, this subset defines the final visually inspected sample distributed across the $1~\mathrm{deg}^{2}$ field (Fig.~\ref{fig:spatial_distribution}). Their SEDs display a range of morphologies, from warm excesses emerging at $\sim$3.6~$\mu$m to more gradual rises at longer wavelengths, consistent with cooler or more extended emitting components. Representative SEDs are shown in Fig.~\ref{fig:sed}.

Time-series analysis of the 446 infrared-excess sources, carried out as
described in Section~\ref{sec:var}, yielded Stetson indices above the adopted variability threshold
for 356 sources. After applying the periodicity and visual-inspection criteria described in Section~\ref{sec:var}, the final
sample comprises 171 variable sources. The detected periods range from days to
multi-year timescales, indicating a heterogeneous set of variability
mechanisms, potentially including stellar rotation, eclipses, or
disk-related obscuration events. Examples of light curves are
presented in Fig.~\ref{fig:lightcurve}.

The distribution of distances for the 171 infrared-excess variable sources is
shown in Fig.~\ref{fig:distance}, indicating that the sample spans a broad
range of distances within the Galactic disk. The effective temperature
distribution, derived from SED fitting and presented in Fig.~\ref{fig:teff},
reflects the diversity of stellar types within the main-sequence candidate sample.

Using the best-fitting effective temperatures from VOSA, we assigned approximate spectral types to the 446 infrared-excess sources using the $T_{\rm eff}$--spectral type calibration of \citet{PecautMamajek2013}. This classification is intended only as a photometric consistency check. The fitted surface gravities should also be interpreted with caution, as broad-band SEDs alone do not strongly constrain $\log g$ for all sources. We therefore use the fitted $T_{\rm eff}$ values primarily to assign approximate spectral classes, while treating discrepant or poorly constrained $\log g$ values as indicators of possible contamination or uncertain SED fits.

To assess the reliability of the detected variability, the sample of variable
infrared--excess sources was cross-matched with external catalogues of known
variables, including the \textit{Gaia} DR3 variability catalogue
\citep{Eyer2023}, the VIVACE classification ensemble \citep{vivace}, and the
VIVA-I infrared variability catalogue \citep{vivai}, adopting positional
tolerances consistent with the astrometric uncertainties of each survey.
We identify counterparts previously classified as variable sources, encompassing
pulsating variables ($\delta$~Scuti/$\gamma$~Doradus/SX~Phoenicis), eclipsing
binaries, RS~CVn systems, and solar-like variables (Table~\ref{tab:known_variables}).
For sources with independently reported periods in the VIVA-I catalogue, the
published values are consistent with those derived from our light-curve
analysis. Sources without counterparts in existing catalogues remain new
variability candidates and constitute promising targets for future follow-up
observations.

The full infrared-excess catalogue is provided as online supplementary material in machine-readable format. A representative subset is provided in Appendix~A, including stellar parameters from SED fitting and variability properties derived from the VIRAC2 light curves.

\begin{table*}
\centering
\scriptsize
\setlength{\tabcolsep}{4pt}
\caption{Infrared--excess sources in our sample with counterparts previously classified as variables in external catalogues.}
\label{tab:known_variables}
\begin{tabular}{cccccc}
\hline
VIRAC2 ID & RA (deg) & Dec (deg) & Catalogue & Variability Class & Period \\
\hline
15983390002907 & 176.39679 & -60.92202 & Gaia DR3 & DSCT/GDOR/SXPHE & --- \\
15995944001829 & 176.84320 & -61.12437 & Gaia DR3 & DSCT/GDOR/SXPHE & --- \\
15998449003218 & 177.95508 & -61.18936 & Gaia DR3 & DSCT/GDOR/SXPHE & --- \\
16000935002314 & 176.80647 & -61.23682 & Gaia DR3 & DSCT/GDOR/SXPHE & --- \\
16003429001826 & 177.38278 & -61.32274 & Gaia DR3 & DSCT/GDOR/SXPHE & --- \\
16013354000688 & 178.01595 & -61.48977 & Gaia DR3 & DSCT/GDOR/SXPHE & --- \\
16018280001452 & 176.60733 & -61.59341 & Gaia DR3 & DSCT/GDOR/SXPHE & --- \\
16018282000324 & 176.90109 & -61.60201 & Gaia DR3 & DSCT/GDOR/SXPHE & --- \\
16025651003078 & 176.06307 & -61.69538 & Gaia DR3 & DSCT/GDOR/SXPHE & --- \\
16025654000071 & 176.54443 & -61.73043 & Gaia DR3 & DSCT/GDOR/SXPHE & --- \\
16035432000469 & 176.77044 & -61.86468 & Gaia DR3 & DSCT/GDOR/SXPHE & --- \\
15998440001680 & 176.67160 & -61.22240 & Gaia DR3, VIVACE & ECL / EA-EB & --- \\
16000934000409 & 176.56275 & -61.23114 & Gaia DR3, VIVACE, VIVA-I & ECL / EW & $P_{\rm VIVA-I}=0.3753$ d \\
15990935002808 & 176.73245 & -61.01834 & Gaia DR3 & RS CVn & --- \\
16013342001900 & 176.21798 & -61.46864 & Gaia DR3 & RS CVn & --- \\
16015819002744 & 177.26263 & -61.51031 & Gaia DR3 & RS CVn & --- \\
16035429001276 & 176.32427 & -61.87429 & Gaia DR3 & RS CVn & --- \\
15995946000529 & 177.24234 & -61.15676 & Gaia DR3 & Solar-like & --- \\
15995947002537 & 177.37698 & -61.13503 & Gaia DR3 & Solar-like & --- \\
16003422000543 & 176.25197 & -61.28125 & Gaia DR3 & Solar-like & --- \\
15993447001803 & 177.61858 & -61.09267 & VIVA-I & Infrared variable & $P_{\rm VIVA-I}=0.0480$ d \\
16018289002221 & 177.94916 & -61.57311 & VIVA-I & Infrared variable & $P_{\rm VIVA-I}=0.0572$ d \\
\hline
\multicolumn{6}{l}{\footnotesize Notes. DSCT/GDOR/SXPHE denotes $\delta$~Scuti/$\gamma$~Doradus/SX~Phoenicis candidates; ECL denotes eclipsing binaries.}
\end{tabular}
\end{table*}

\section{Discussion}
\label{sec:discussion}

The 446 infrared-excess sources identified in the VVV tile d077 represent a population that has been largely inaccessible in previous infrared-excess and debris-disk searches, which typically avoid the Galactic plane due to high extinction and source crowding. The combination of deep optical, near-infrared, and mid-infrared photometry used here demonstrates that multi-survey datasets can overcome these limitations and enable systematic searches for infrared-excess sources in highly reddened regions.

The classification of the selected sources as main-sequence candidates is based on empirical photometric criteria and should therefore be interpreted with caution. The SED-derived effective temperatures provide an independent consistency check and allow approximate spectral types to be assigned, but broad-band photometry alone cannot uniquely distinguish all dwarfs from reddened giants, evolved stars, young stellar objects, or blended sources. Consequently, a fraction of the infrared-excess catalogue may still include contaminants. This limitation does not affect the usefulness of the catalogue as a candidate list, but it reinforces the need for spectroscopic validation before individual sources can be confirmed as debris-disk systems.

The diversity of SED morphologies in our sample suggests a range of possible physical scenarios, including different dust temperatures, disk geometries, and evolutionary stages. Sources whose excess emerges at shorter infrared wavelengths (e.g., $3.6~\mu$m) are consistent with the presence of relatively warm dust, while those with excess confined to longer wavelengths may trace cooler and more extended dust distributions. However, these interpretations are not unique, as unresolved binary companions or line-of-sight blends may also contribute to the observed infrared-excess, introducing degeneracies in the physical interpretation.

The detection of periodic variability in 171 of the excess sources shows that time-domain information provides a valuable complement to SED-based diagnostics. While some periodic signals may be associated with stellar rotation, others could be linked to occultations by circumstellar material or transient dust structures. The coexistence of infrared-excess and coherent variability suggests that a subset of these systems may host dynamically evolving environments. Nevertheless, the relatively high incidence of variability should be interpreted with caution, as a direct comparison with field-star variability fractions requires a control sample selected with consistent photometric and temporal criteria. In addition, unresolved binaries and blending effects may contribute to the observed variability.

A key limitation affecting the interpretation of these results is the presence of unresolved binary systems and line-of-sight blending. A substantial fraction of field stars are members of binary or multiple systems, with estimates exceeding 50\% for solar-type stars \citep{DuquennoyMayor1991}. In unresolved systems, the combined flux of multiple components can mimic infrared-excess, potentially contaminating debris-disk samples. Similarly, source crowding in the Galactic plane can lead to artificial excesses due to blending. While our selection procedure reduces these effects, it cannot fully eliminate them.

Consistent with these limitations, a small subset of sources appears fainter than the main sequence while still exhibiting infrared-excess. These cases may be explained by binarity, extinction effects, or photometric uncertainties, all of which can alter the observed flux distribution and produce apparent excesses in colour--magnitude space.

A more quantitative assessment of these effects requires spectroscopic observations capable of identifying radial-velocity variability and composite spectral signatures. Therefore, the infrared-excess sample presented here should be interpreted as a catalogue of candidates rather than as a confirmed debris-disk population.

Future spectroscopic follow-up is essential to establish the nature of the detected excesses. Near-infrared spectroscopy can provide independent constraints on stellar parameters and enable the identification of spectroscopic binaries. Multi-object facilities such as KMOS, particularly within the context of surveys like VVVX-GalCen (e.g., Gomez et al., in preparation), are well suited for this purpose, while upcoming facilities such as MOONS will further enhance the efficiency of large-sample characterization. In addition, mid-infrared spectroscopy with JWST/MIRI may constrain dust composition, and high-resolution (sub-)millimeter observations with ALMA can probe cold dust components and disk structure.

\section{Conclusion}

We have presented a multi-wavelength method designed to identify main-sequence candidates with infrared excess in crowded and high-extinction regions of the Galactic plane. By combining optical, near-infrared, and mid-infrared photometry from DECaPS, VIRAC2, and GLIMPSE, we constructed reliable spectral energy distributions (SEDs) for 3,169 sources in VVV tile d077 and identified 446 sources flagged as having infrared excess. These excesses are consistent with the presence of circumstellar dust, although contributions from unresolved binary companions cannot be excluded. Time-series analysis of VIRAC2 $K_{\rm s}$-band data further revealed 171 sources with coherent periodic variability, highlighting a subset of systems in which dust dynamics, occulting structures, or other astrophysical processes may play an important role.

These results demonstrate that infrared-excess candidates can be systematically identified even in the challenging environment of the inner Milky Way. The observed diversity in excess morphologies and variability behaviour indicates a heterogeneous population, likely influenced by a combination of circumstellar dust, stellar properties, and binary-related effects.

These findings motivate further follow-up studies. Spectroscopic observations will not only refine stellar parameters and help identify unresolved binaries, but also provide an independent validation of the photometric target-selection and classification methodology adopted here. Such observations will be essential to verify whether the selected sources have spectral types consistent with the SED-based classifications and to quantify the contamination rate from non-main-sequence stars or blended systems. High-resolution observations with ALMA can probe cold dust components and disk structures. Moreover, the scalability of our approach enables this analysis to be extended to the full VVV survey. Future improvements, including automated main-sequence classification, proper-motion propagation, and improved deblending techniques, will further enhance this potential.

The catalogue therefore provides a robust starting point for future studies aimed at confirming and characterising the nature of circumstellar environments in the Galactic plane, particularly in regions that have been historically underexplored.

\section*{Acknowledgements}

P.E. acknowledges the VVV collaboration for their support and for fostering a collaborative research environment. P.E. also acknowledges support from Conselho Nacional de Desenvolvimento Científico e Tecnológico (CNPq) through an undergraduate research fellowship. C.C. acknowledges support from ANID BASAL project FB210003. D.M. acknowledges support from the BASAL Center for Astrophysics and Associated Technologies (CATA) through ANID grants ACE210002 and AFB 210003, and from Fondecyt Regular project 1220724. Z.G. acknowledges support from the China-Chile Joint Research Fund (CCJRF No. 2301) and the Chinese Academy of Sciences South America Center for Astronomy (CASSACA) Key Research Project E52H540301. Z.G. is also supported by the National Research and Development Agency (ANID) through the FONDECYT Iniciación project Grant No. 11260176. R.K.S. acknowledges support from CNPq/Brazil through projects 308298/2022-5 and 421034/2023-8. This publication makes use of VOSA, developed under the Spanish Virtual Observatory project supported by the Spanish Ministry of Science and Innovation through grant AyA2017-84089. D.Q. acknowledges financial support from the Brazilian Federal Agency for Support and Evaluation of Graduate Education – CAPES – Finance Code 001.

\section*{Data Availability}

The data underlying this article are available in the cited public surveys. The full machine-readable catalogue of infrared-excess sources presented in this work is available as online supplementary material. A representative subset is provided in Appendix~A.

\bibliographystyle{mnras}
\bibliography{references}

\appendix

\section{Representative subset of the infrared-excess catalogue}

A representative subset of the infrared-excess catalogue is shown in Table~\ref{tab:sample_catalog};
the full machine-readable version is described in the Data Availability section.

\begin{table*}
\centering
\caption{Representative subset of the infrared-excess catalogue, including identifiers, astrometry, best-fit stellar parameters from SED fitting, extinction, distance, and variability diagnostics. The full machine-readable table is available as online supplementary material.}
\label{tab:sample_catalog}

\resizebox{\textwidth}{!}{%
\begin{tabular}{ccccccccccccccc}
\hline
VIRAC2 ID & Gaia DR3 ID & RA & Dec & Model & $T_{\rm eff}$ & Approx. SpT & $\log g$ & $A_V$ & D (pc) & Stetson & FAP $< 1\%$? & $P_{\rm LS}$ (d) & Best Power & $P_{\rm BLS}$ (d) \\
\hline
15983390002907 & 5335051458521230464 & 176.39678792 & -60.92201599 & BT-Settl-agss & 11200.0 & B9 & 3.0 & 4.232 & 4511 & 0.69103 & False & 0.044393 & 11.092493 & 0.227244 \\
15983391002028 & 5335050943125213440 & 176.51215859 & -60.92508884 & BT-Settl-agss & 5700.0 & G3 & -0.5 & 1.096 & 2237 & 0.4421 & False & 0.052032 & 115.042656 & 1.722249 \\
15983391002835 & 5335051046204484608 & 176.55413024 & -60.91136767 & BT-Settl-agss & 6200.0 & F8 & -0.5 & 0.568 & 1415 & 0.60645 & False & 0.061256 & 11.549294 & 0.874149 \\
15983392000057 & 5335005450826508160 & 176.64788121 & -60.90328033 & BT-Settl-agss & 5600.0 & G5 & -0.5 & 0.575 & 4319 & 0.45052 & False & 0.046754 & 11.272844 & 1.505373 \\
15983392001939 & 5335005553905703680 & 176.63251933 & -60.91514252 & BT-Settl-agss & 9200.0 & A1 & 0.5 & 2.683 & 3668 & 0.67435 & False & 0.063185 & 8.621822 & 0.692063 \\
15983393001227 & 5335005038509666816 & 176.72661683 & -60.90355129 & BT-Settl-agss & 6000.0 & G0 & -0.5 & 1.540 & 5237 & 0.43533 & False & 0.043126 & 9.882275 & 0.285065 \\
15983394002193 & 5335377154470733824 & 176.89294877 & -60.91727413 & Kurucz2003 & 9500.0 & A0 & 3.5 & 1.363 & 2055 & 0.45727 & False & 0.065391 & 6.76624 & 0.315837 \\
15985908000005 & 5335045445566640896 & 176.37695373 & -60.98286483 & BT-Settl-agss & 6400.0 & F6 & -0.5 & 0.890 & 1754 & 0.62884 & True & 0.053244 & 8.491941 & 0.210878 \\
15985909001621 & 5335050633891547904 & 176.47069204 & -60.95164985 & BT-Settl-agss & 4300.0 & K5 & 5.0 & --- & 3945 & 0.59386 & False & 0.129492 & 5.789726 & 0.340141 \\
15985910004574 & 5335003904638212352 & 176.63157367 & -60.95854076 & BT-Settl-agss & 8400.0 & A4 & 1.0 & 3.574 & 4920 & 0.7206 & False & 0.117929 & 7.238211 & 0.290063 \\
15985911000998 & 5335004488753797248 & 176.71100047 & -60.94632794 & BT-Settl-agss & 5900.0 & G1 & -0.5 & 0.830 & 2048 & 0.72982 & False & 0.068869 & 9.91579 & 0.539083 \\
15985911001377 & 5335004626192781824 & 176.72439840 & -60.92505019 & BT-Settl-agss & 5400.0 & G8 & -0.5 & 0.354 & 2136 & 0.75883 & False & 0.049585 & 9.652795 & 0.951374 \\
15985912002402 & 5335001705615332992 & 176.90579909 & -60.95968505 & BT-Settl-agss & 9600.0 & A0 & 1.5 & 1.913 & 2880 & 0.44767 & False & 0.045472 & 11.133528 & 0.437652 \\
15985913001285 & 5335375848800654976 & 177.09811775 & -60.98491845 & BT-Settl-agss & 9000.0 & A2 & 3.0 & 1.068 & 2056 & 0.37049 & False & 0.376531 & 23.440757 & 1.000080 \\
15985913002387 & 5335376123678566016 & 177.01608251 & -60.95956308 & BT-Settl-agss & 4600.0 & K4 & 5.5 & --- & 3308 & 0.59824 & False & 0.117289 & 5.574945 & 0.266641 \\
15988422001690 & 5335045239408206592 & 176.33018159 & -60.99283235 & BT-Settl-agss & 5100.0 & K1 & -0.5 & --- & 4983 & 0.68435 & False & 0.117914 & 9.02801 & 0.233810 \\
15988422002103 & 5335044925818165248 & 176.36371111 & -61.01673040 & BT-Settl-agss & 5600.0 & G5 & -0.5 & 0.593 & 1652 & 0.75481 & True & 0.162515 & 10.345977 & 0.360918 \\
15988423002048 & 5335044380414571904 & 176.49800372 & -61.01598392 & BT-Settl-agss & 5000.0 & K2 & 5.5 & 0.141 & 1257 & 0.63357 & False & 0.273788 & 7.999968 & 0.244982 \\
15988423002143 & 5335044616580603904 & 176.41956048 & -60.99275566 & BT-Settl-agss & 7800.0 & A7 & 0.5 & 3.381 & 7045 & 0.4796 & True & 0.052580 & 11.777042 & 0.550746 \\
15988424000486 & 5335003354882348032 & 176.60902437 & -61.00227659 & BT-Settl-agss & 5900.0 & G1 & 0.5 & 0.414 & 806 & 0.55918 & True & 0.042216 & 9.515991 & 0.547315 \\
\hline
\end{tabular}%
}
\parbox{0.95\linewidth}{\footnotesize Notes. Approximate spectral types are inferred from the best-fitting $T_{\rm eff}$ values and should be interpreted as photometric, not spectroscopic, classifications.}

\end{table*}

\end{document}